\documentclass[aps,pre,preprint,nofootinbib,floatfix, showpacs]{revtex4}
\usepackage{graphicx}
\usepackage{amsmath}
\bibpunct[, ]{[}{]}{,}{a}{,}{,}
\newcommand{\be}{\begin{equation}}
\newcommand{\ee}{\end{equation}}

\begin{document}
\title{Physical Schemata Underlying Biological Pattern Formation - Examples, Issues and Strategies}

\author{Herbert Levine}
\affiliation{Center for Theoretical Biological Physics, University of California at
San Diego, La Jolla, CA 92093-0319 USA}
\author{Eshel Ben-Jacob}
\affiliation{School of Physics and Astronomy.  Raymond and Beverly Sackler Faculty of Exact Sciences, Tel-Aviv Univ.,69978, Tel-Aviv, Israel}
\date{\today}
\begin{abstract}
Biological systems excel at building spatial structures on scales ranging from nanometers to kilometers and exhibit temporal patterning from milliseconds to years. One approach that nature has taken to accomplish this relies on the harnessing of pattern-forming processes of non-equilibrium physics and chemistry. For these systems, the study of biological pattern formation starts with placing a biological phenomenon of interest within the context of the proper pattern-formation schema and then focusing on the ways in which control is exerted to adapt the pattern to the needs of the organism. This approach is illustrated by several examples, notably bacterial colonies (diffusive-growth schema) and intracellular calcium waves (excitable-media schema).
\end{abstract}
\maketitle
\section{Introduction}

Biological pattern formation, by which we mean the organization of living matter on scales ranging from nanometers to kilometers, is a vast multifaceted area of intense, ongo-ing research. Similarly, there is a large community of scientists working towards a better understanding of patterns in myriad abiotic systems, where much progress has been made in identifying basic physical mechanisms (ÒschemataÓ) underlying the diverse observations. To give just one example, short-range activation and long-range inhibition underlies periodic patterns of spots and stripes (ÒTuringÓ structures). It is therefore natural to wonder as to the overlap of these two subjects, at the issue of the extent to which pattern-formation physics is an important ingredient for understanding pattern-formation biology.

In our opinion, there are both clear examples and clear counter-examples of this rele-vance. On the positive side lie a vast array of microorganism colony structures (in Bacillus and related species [1], in myxobacteria [2], and in eukaryotes such as Dictyostelium [3]) and intracellular waves [4]; other unproven possibilities include activity-dependent structures in the developing nervous system, tropical fish coloration, and seashell patterns. Here, we will use several of these systems to offer a perspective on this area of mutual physics/biology interest. But it is important to keep in mind that what we say probably will not apply to many other processes. Perhaps the clearest such negative instance is afforded by stripe for-mation during early Drosophila embryogenesis. Early attempts to identify stripe formation with the Turing pattern mechanism foundered upon the demonstration that each stripe is separately controlled by dedicated transcription regulation. This system is more akin to building an office tower from blueprints than is to the beautiful Turing structures found in nonlinear chemical systems. This does not mean that physics is irrelevant (any more than it is for the building) but the biological pattern is not built upon a physical pattern schema.

Our purpose here is two-fold. For the pattern-formation physicist, we want to explain the challenges that must be addressed upon making the transition from studying non-living to living systems. Insufficient attention to the underlying purpose of the patterning phenomena can lead one to miss the some of the most interesting aspects of these systems, namely how the physics is sculpted by a plethora of feedback mechanisms and how degrees of freedom at different scales can be strongly coupled. For the biologist, we want to advance the claim that it will often be necessary to work out the details of the physical basis underlying a specific biological pattern. Just as evolution can create genes that directly encode for structure, it can create genes that utilize and guide non-equilibrium physical processes to produce adaptive patterns.
\section{Bacterial Colony Patterns}
\subsection{Introduction}

Bacterial colonies are a natural meeting ground for the physics and biology of pattern formation. Traditionally, bacterial colonies in the lab are grown on substrates with a high nutrient level and intermediate agar concentration. Such  ÒfriendlyÓ conditions yield colonies of simple compact patterns that fit well the old-fashioned view of bacterial colonies as a collection of independent unicellular organisms (non-interacting ÒparticlesÓ). However, bacterial colonies in nature must regularly cope with hostile environmental conditions and it is reasonable to guess that the colonies must be prepared to exhibit complex patterns which serve as an adaptive response. Mimicking this by the creation of hostile conditions in a Petri dish leads to an incredible diversity of structures (Fig. 1) that pose multiple challenges for the physical scientist interested in discerning the underlying mechanisms and in making sense of all this rich behavior.

The first step is to identify the pattern formation schemata upon which we can begin to build models and make predictions. The first patterns to be carefully studied were branched fractal structures highly reminiscent of those seen in phys-ical systems under conditions of diffusion-limited growth [5]. Diffusion-limited growth is the name associated with a variety of physical and chemical processes in which one phase of matter displaces another, but at a rate that is limited by the diffusive transport of some needed substance. The most familiar example is probably crystal growth from supersaturated solution in which the solid phase grows by the attachment of molecules diffusing through the liquid. The diffusion field drives the system towards decorated (on many length scales) irregular fractal shapes, due to the action of the Mullins-Sekerka (MS) instability: outward protrusions are more effective at attracting flux and therefore grow faster than their surroundings. 

This identification seems quite straightforward. But looks can be deceiving and one needs to devise both rigorous tests of this hypothesis and plausible biological underpinnings for the diffusive instability; only afterwards can we delve more deeply in the specifically biological issues. For the former, we point to Fig 2. where the prediction that small-scale external variations in growth conditions can lead to macroscopic re-organization of the pattern is verified in a {\it Paenibacillus dendritiformis} colony. As to the biology, it has been shown (using a variety of models) that the need for nutrient, coupled to having a sharp colony edge, will indeed lead to an MS instability. In the original experiments of Matsushita et al [6], the sharpness is accomplished by making the surface so hard that the bacterial strain ({\it Bacillus subtilis}) cannot swim across the surface; motion of the colony occurs by individual cell lengthening followed by cell division in situ at the edge of the colony. The newly identified {\it Paenibacillus dendritiformis} bacteria use an elegant ÒtrickÓ to accomplish the same thing. The bacteria emit lubricating fluid to enable themselves to move on the substrate; they adjust the time constant of lubricant production, its viscosity, and its absorption rate back to the substrate such that the effective bacterial diffusion constant will increase (from zero) with increasing bacterial density; this is called nonlinear diffusion. In addition, bacteria further back (where the food level is lower) enter into a pre-spore state in which they are non-motile (but still maintain chemical communication, as we will describe latter).  The combined effect of non-linear diffusion and pre-sporulation leads to the growth of branches with well-defined boundaries. This mechanism can be explicitly incorporated in a reaction-diffusion framework [7], thereby leading to a physics-based model for the branching pattern.

\subsection{Beyond branching physics}

Placing the bacterial colony systems in the diffusive growth schema enables us to get started with the task of understanding and modeling the observed dynamics. However, it is crucial to realize that inherent in the growth of a colony is an additional degree of plasticity: the building blocks of the colony are themselves living organisms, each with internal degrees of freedom, internally stored information and an internal assessment of external chemical messages. These afford each bacterium freedom to respond flexibly and even alter itself, by means of modifying its genetic expression pattern. In fact, we have already seen that flexibility when the bacteria use lubricating fluid to create the conditions responsible for branching. We mention in passing the established fact that the branching patterns can be modified by further bacterial cooperativity, typically through the exchange of both attractive and repulsive chemotactic (i.e. motion-biasing) signals.

     How much control can the bacteria exert? We have already seen in Fig 2. evidence of the still surprising finding that diffusion-limited patterns can be controlled by external ma-nipulation at the ÒmicroscaleÓ. But what can be externally controlled can also be internally controlled, i.e. can be manipulated by changing the local interactions between the bacteria. This is the secret behind a totally different morphotype in which all the branches are twisted with the same handedness (Fig 1b). These patterns are called chiral branching (CB) and appear to result from the amplification of the intrinsic chirality of the bacterial flagellum. The bac-teria effect this transition by suppressing cell division, leading to a typical cell length of five to ten times the length in the branching state. This changes the mode of cell motion from the typical run and tumble of the short cell to a quasi-one-dimensional forward-backward movement with a chiral bias for the CB morphotype bacteria. Simulation studies (see Fig. 3) using models that impart to the bacteria an extra orientational degree of freedom have verified that these microscopic changes can indeed alter the global pattern.

We do not as yet understand the actual mechanism underlying the change at the single bacterium level. Since the changes are heritable, the most likely possibility is a master switch in the gene regulation network. Such switches clearly exist and are responsible for cell differentiation, as seen for example during sporulation; in sporulation, flipping the switch requires messages from other cells and hence implements a sort of collective com-putation of the colony-scale response appropriate to a given set of external conditions. This raises the important question of whether the decision to become elongated is made individually or collectively, is made deterministically or randomly. Given the run-to-run vari-ation in when (and whether) transitions occur, there is clearly some element of randomness involved. On the other hand, modeling suggests that we need a finite-sized group of bacteria to initiate a chiral branch from the ordinary (wider) branch, similar to the familiar need for a finite bubble size of nucleation in crystal growth. The only two possibilities are either the type of cell-to-cell communication just discussed or an as-yet-unknown mechanism for initially sparse populations of individual elongated cells to find each other. There is therefore an urgent need for experiments that can elucidate the actual molecular biology of this remarkable phenomenon.

\subsection{Additional levels of organization}

We have been talking until now about patterns that involve cellular degrees of freedom that interact to directly form structures at the colony scale. But there is another option, namely that the bacteria form mesoscale structures which then become building blocks used to form the colony. Consider the patterns formed by the strain {\it Paenibacillus vortex}. The bright spots at the end of the branches are rotating vortices of bacteria which collectively move out across the agar, leaving behind trails of cells (see Fig 2f). Once formed, the vortex expands and translocates as a unit, leaving behind a trail of cells forming its own branch. Modeling has helped show that a self-generated, short-range chemoattractant can be utilized for vortex formation while a self-generated, long-range chemorepellent can push the vortices outwards, thus bringing about the expansion of the colony.

The vortex pattern seems to illustrate another general principle of biological systems, that they can be organized hierarchically. Once the vortex is recognized as a possible spatial structure, it becomes easy to see that vortices can be used as subunits in a more complex colonial organization. In fact one can go a step higher; once a colony is recognized as a possible spatial structure, one should expect that colonies can be used as building blocks of even more complex structures. This indeed occurs in biofilms, where individual colonies communicate so as to better accomplish necessary tasks, even going so far as to exchange relevant genetic information.
\subsection{ Adaptive Utility of the patterns}

One typical reaction to these patterns is that they are very ÒartisticÓ but could not possibly have any role in the survival of the organism. This opinion seems to be based on the notion that only dynamics directly encoded by the genes could be acted upon by selection, and that the existence of a physical schema underlying the system clearly ÒprovesÓ that the pattern is merely an incidental consequence of the physics of the bacterial system. We believe otherwise. First, the bacteria devote a great deal of effort to modulate their behavior to enable patterning; recall the lubricating fluid trick, the turning on and off of chemotactic signaling, and the altered state responsible for CB. The patterns are highly reproducible at the ÒmacroscaleÓ, even in the presence of notoriously noisy cell-scale dynamics; of course it is this overall organization of the pattern, not the microscopic details, which could be the object for selection. The resultant structures appear adaptive: branching allows more effective spread across the nutrient-poor surface; (attractive) chemotactic signaling has been suggested as a strategy in the response to oxidative stress [8]; and data to date suggests that the bias between B and CB structures is determined by which one moves the fastest. Recent experiments using sub-lethal levels of antibiotic have attempted to directly es-tablish the functional utility of pattern variations. In Fig 4., we show close-up pictures of the effect of non-lethal levels of septrine on the branch structure of the {\it Paenibacillus dendritiformis} CB morphotype. Simulations suggest that the antibiotic causes the system to amplify repulsive signaling so as enable the colony to more effectively Òrun awayÓ from regions known to contain antibiotic.

\subsection{Reprise}

In studying the bacterial colony, we have learned several lessons regarding how the bacteria turn the physics of diffusive patterning into a functionally important part of their survival strategy. We believe that these lessons are in fact generally useful in thinking about this class of biological pattern-forming dynamics. To illustrate our thinking, we will now use this perspective to discuss intracellular calcium patterns.

\section{Intracellular Calcium Signaling}

\subsection{Introduction}

Consider the dynamics of calcium inside eukaryotic cells [9]. The basic phenomenon is the propagation of a non-linear chemical wave, supported by an amplification mechanism known as calcium-induced calcium release (CICR). To understand this, we must recognize that a typical cell contains many internal compartments, each with its own nearly impermeable membrane. Inside some of these, notably the endoplasmic reticulum and the mitochondria, there are high resting calcium concentrations compared to those in the cytosol. CICR refers to the fact that cytosolic calcium can act to open ligand-gated calcium channels, converting a small initial perturbation into a large pulse via gradient-driven onrush. This local increase in calcium can then spread to other points in the cell, without any diminution of amplitude. This phenomenology places the system under the excitable-media schema; well-studied examples of excitable media range from the Belousov-Zhabotinsky chemical reaction to the chemistry of carbon monoxide catalysis on metallic surfaces (the Òcatalytic converterÓ) and to electrical activity in heart tissue consisting of cells whose membranes contain voltage-gated ion channels.

As before, we must first verify the schema identification. Evidenced gathered to date in cells such as the Xenopus oocyte leaves no doubt that calcium dynamics can be approached via the excitable medium conceptual framework [10]. For example, we see the prototypical rotating spiral in Fig. 5a. For these waves: wavefronts clearly annihilate [11]; there is clearly a refractory period for subsequent excitation; and propagation clearly fails when the threshold for excitation (i.e. for the opening of a cluster of calcium channels) is too high. All of these are expected properties of excitable waves. Interestingly, there is some new physics here, as the small number of channels leads to the importance of stochasticity in the wave dynamics [12].

\subsection{Control of the patterns}

The first lesson from the bacteria schemata is that biological systems excel at controlling patterns via coordinating and/or modifying the underlying constituents. This allows the basic system to be flexibly adapted for a variety of purposes. What are the degrees of freedom available to control the nature of intracellular calcium waves? Possibilities include the activities and spatial arrangements of the underlying protein complexes such as the receptors and pumps, the complement of buffers in the cytosol, and the geometry of the cell itself. With these options, the task at hand becomes that of identifying how these elements could in fact alter the patterns and that of studying the experimental signatures of each of these alterations. As we will see, there are many clues that the patterns are being controlled, but much more work is needed to pin down the details and the implications.

${\boldmath IP_3} $ {\bf  control:} The simplest example concerns the use of second messengers such as $IP_3$ to affect the behavior of the $IP_3R$ calcium channel. In some experiments [13] $IP_3$ is held (roughly) constant. This work has shown that the CICR mechanism on its own is sufficient for (at the least noisy) oscillatory behavior and that increasing $IP_3$ corresponds to increasing the excitability and, to a lesser extent, decreasing the stochasticity of the system. This excitability increase causes a transition from an isolated ``puff" state to abortive waves and finally to propagating regenerative waves (Fig. 5b-d). Within the propagating wave range,  the average period between successive global waves (events where a set of nearby puffs nucleate a wave which passes through much of the cell  decreases as a function of increasing $IP_3$. 

What happens if $IP_3$ is not artificially controlled, i.e. if we allow the cell to engineer its own waves via unconstrained $IP_3$ dynamics? In a variety of cell types, $IP_3$ will oscillate along with the calcium concentration but it is often unclear as to whether this is merely a side-effect of the CICR-driven dynamics or whether $IP_3$ dynamics is intrinsically critical. From our perspective, the most interesting variant of this question is whether the feedback coupling calcium cycling to lipid metabolism can give the system more flexibility in responding to stimuli. One idea along these lines that has arisen both from experiments [14] and models is that the coupled oscillations can be of a more complex, bursting character; exactly what additional information about external conditions could be encoded in this manner is unfortunately not clear.

{\bf Cell organization:} Even with fixed components, changes in the organization of the cell can alter the particular form that calcium signals will take. For example, the exact number of channels that make up each cluster as well as the spatial pattern of the clusters can have large effects. It has been suggested [15] that at least in the oocyte these are chosen in some optimal way to generate robust oscillatory responses. Too large a cluster spacing with too many channels per cluster (so as to maintain a fixed total number of channels) cannot lead to either waves or global oscillations; no clustering at all leads to very noisy behavior, as there is no averaging away of any of the inherent stochasticity of the single channel. If this is true, it immediately raises the question of the mechanism enforcing this optimality. Is this configuration somehow encoded directly in the details of the genes (as realized in the observed membrane-protein interactions) or is there feedback from the calcium dynamics to these other levels, perhaps even with real-time adjustments? Some hints that this type of feedback may be possible arise from studies of the effects of calcium on the clustering dynamics [16]; however, as with most of the other issues in this section, more work is needed to provide a definitive answer.

Similar phenomena occur upon varying calcium buffers. Cells can change their calcium-handling systems in response to external signals. Thus the maturation of the oocyte is accompanied by a change from calcium pulses to a calcium front by which a high-concen-tration state supplants the usual low concentration. This behavior is referred to as bistable and models of the fertilization wave adjust the various terms in the kinetic equations to assure bistability. This of course begs the question of how the cell ensures bistability.

\subsection{Higher-level Organization}

What we have tried to stress by this parallel look at intracellular calcium dynamics and bacterial colony organization is that the strategies used by living systems to create functional patterns are used repeatedly in many different contexts. One can therefore generalize from one problem to another, not at the level of specific details but at the more fascinating level of conceptual frameworks. If we continue along this line, we can make several predictions.

{\bf Constituents are never simple and rarely act alone:}. Bacteria can communicate so as to alter the patterns. We thus expect that the protein machines involved in calcium handling will prove to be much more sophisticated than they are given credit for today. In the context of bacterial chemotaxis, the idea that receptors are fixed molecules which act as lone sensors has given way to a much richer view of dynamically interacting receptor assemblies in which each receptor can be in a wide range of different states (differing by methylation of residues, e.g.). We can thus speculate that detailed investigations will reveal links between receptors (some evidence of this has already been claimed in skeletal muscle [17] and cardiac cells), and also will reveal important internal degrees of freedom of the receptor molecule.

{\bf Systems can be used as subsystems to create more complex behavior:}  Once one understands that the cell can act as the ``macroscale" for the self-organized dynamics of the calcium processing ``microscopic" dynamics, it is not surprising that cells themselves can interact via {\em intercellular} calcium waves, thereby creating a three-level hierarchical pattern. An example of this occurs in the glial network in the brain [18].

There is no real doubt that specific features of these patterns are playing important func-tional roles in a variety of cell types. In the cardiac system, sparks and abortive waves ensure the graded contractile response to cell stimulation. In oocytes and hepatocytes, global waves coordinate gene expression patterns, sometimes in a frequency-specific man-ner. Neural growth cones appear to use calcium pulses to encode information about the current micro-environment.  Most recently [19], rotating pulses in neutrophils, generated by exposure to chemoattractant, appear to encode directional information needed to execute directed motion. We do not understand the codes, but these spatiotemporal patterns are not incidental to the information being conveyed. Finally, we cannot help but mention recent ideas regarding how the pattern of mitochondria inside cells (which after all can be thought of as an intracellular colony) appears to affect the pattern of calcium signaling - see Fig. 6; perhaps our two topics as not as disjoint as they seem.

\section{Summary and Outlook}

As we have argued throughout, understanding pattern formation in living systems requires a sophisticated dialogue between physics and biology. The physics of non-equilibrium systems sets the playing field upon which the biological processes can act. We have seen explicitly how this dialogue is unfolding for two different example, bacterial colonies shaped by diffusive instabilities and calcium waves governed by nonlinear amplification during intracellular signaling. 

We would like to conclude this perspective/mini-review by presenting a series of lessons that the two of us have had to learn to be able to work in this field. How general these will turn out to be is anyone's guess, but we proceed nonetheless.
\begin{itemize}
\item ¥ Biological systems evolved to survive under trying circumstances. Thus, laboratory protocols must place systems under a variety of stresses in order to explore the func-tional capabilities of these systems.
\item ¥ We will never fully understand biological systems without taking into account the physics and chemistry, which form the underpinning of their capabilities.
\item Conversely, we will never fully understand biological systems without recognizing that  the requirement of flexible functionality guides the use of the aforementioned physics and chemistry.
\item Modeling is somewhat of an art form. The best models focus on a specific set of questions, keep only the pieces that are believed to be essential, and then make testable, nontrivial predictions.

\end{itemize}

When we say that our understanding of a specific pattern-forming mechanism, say bacte-rial colony growth, has advanced, what do we actually mean? It does not mean that we can make quantitatively reliable first-principles calculations of specific patterns in specific instances; we cannot do this even for the physical pattern that started the field, the snowflake grown from supersaturated air. Instead, we have in hand a series of models of increasing sophistication that can make sense of the phenomenology; can reveal to us the key experimental controls that can alter the structures; can allow us to compare and contrast different systems which seem to be in the same conceptual class; and can enable the discovery of new twists built upon the same schemata. With this perspective in mind, it is clear that there will be an exciting future for the study of the biological applications of pattern-formation physics.

The work of HL has been supported in part by the NSF-sponsored Center for 
Theoretical Biological Physics  (grant numbers PHY-0216576
0225630). The work of EBJ was partially supported by the Meguy-Glass chair in Physics of Complex Systems. 

\newpage
\begin{center} REFERENCES \end{center}
\begin{enumerate}

\item Ben-Jacob E, Cohen I and Levine H, ``Cooperative self-organization of microorganisms", Adv. Phys. 49, 395-554 (2000); Ben-Jacob E, ``Bacterial self-organization: co-enhancement of complexification and adaptability in a dynamic environment" Phil. Trans. R. Soc. Lond. A, 361:1283-1312 (2003)

\item Shapiro JA, ``Bacteria as multicellular organisms" Sci.. Am. 258, 62-69, (1988); Losick R and Kaiser D, ``How and why bacteria talk to each other", Cell 73, 873-887 (1993); ``Why and how bacteria communicate" Sci.. Am. 276, 68-73 (1997)

\item Levine H, Aronson I, Tsimring L and Truong TV, ``Positive genetic feedback governs cAMP spiral wave formation in Dictyostelium", PNAS 93, 1151-53 (1996)

\item Berridge MJ and Bootman MD, ``The versatility and universality of calcium signaling", Nat Rev Cell Biol 1, 11-21 (2000)

\item Kessler DA, Koplik J and Levine H, ``Pattern selection in fingered growth phenomena", Adv Phys 37 255-339 (1988); Langer JS, ``Dendrites, viscous fingering, and the theory of pattern formation", Science 243, 1150-54 (1989);  Ben-Jacob E and Garik P, ``The formation of patterns in non-equilibrium growth" Nature 33, 523-30 (1990)

\item Matsushita M and Fujikawa H, ``Diffusion-limited growth in bacterial colony formation" Physica A168, 498-506 (1990)

\item Kozlovsky Y, Cohen I, Golding I and Ben-Jacob E, ``Lubricating bacteria model for branching growth of bacterial colonies" Phys. Rev. E59, 7025-7035 (1999)

\item Budrene EO and Berg HC, ``Dynamics of formation of symmetrical patterns by chemotactic bacteria", Nature 376, 49-53 (1995)

\item Berridge MJ and Bootman MD, ``The versatility and universality of calcium signaling", Nat Rev Cell Biol 1, 11-21 (2000)

\item Sneyd J, Girard S and Clapham D, ``Calcium wave propagation by calcium-induced-calcium release: an unusual excitable system", Bull. Math Biol 55, 315-344 (1993)

\item Lechleiter JD, Girard, S, Peralta, E and Clapham D, ``Spiral calcium wave propagation and annihilation in Xenopus-Laevis oocytes", Science 257, 123-26 (1991)

\item Falcke M, Tsimring L and Levine H, ``Stochastic spreading of intracellular Ca2+ release", Phys Rev E 62 2636-43 (2000)

\item Sun XP, Callamaras N, Marchant JS and Parker I, ``A continuum of InsP$_3$ mediated elementary Ca2+ signalling events in Xenopus oocytes", J Physiol 509, 67-80 (1998)

\item Kummer, U, Olsen LF, Dwon CJ, Green AK, Borberg-Bauer E and Baier G, ``Switching from simple to complex oscillations in calcium signaling", Biophys J 79, 1188-95 (2000)

 \item Shuai JW and Jung P, ``Optimal ion channel clustering for intracellular calcium signaling, PNAS 100, 506-510 (2003)

\item Wilson, BS et al, ``Calcium-dependent clustering of inositol 1,4,5-trisphosphate receptors", Mol Biol Cell, 9, 1465-1478 (1998)

\item Marx SO, Ondrias K, Marks AR, ``Coupled gating between individual skeletal muscle Ca$^{2+}$ release channels", Science 281, 818-21 (1998). 

\item Cornell-Bell AH, Finkbeiner SM, Cooper MS, Smith SJ, ``Glutamate induces calcium waves in cultured astrocytes: long-range glial signaling", Science 247, 470-73 (1990)

\item Kindzelskii, AL and Petty, HR, ``Intracellular Calcium waves accompany neutrophil polarization, FMLP stimulation and phagocytosis: A high speed microscopy study", J. Immunol. 170,64-72 (2003).

\end{enumerate}
\newpage
\begin{center} FIGURE CAPTIONS \end{center}
\begin{trivlist}

\item{Fig 1:} Example of patterns formed during self-organization of lubricating bacterial colonies:a-c are macro-level view of the entire colony, each has about 10$^9$ Ð 10$^{10}$ bacteria. d-e  are the corresponding micro-level view with magnification x500.  Each bar is an individual bacterium.  a,d are of the branching morphotype of the P. Dendritiformis bacteria; note the branching pattern and the random like orientations of the bacteria within the branch.  Yet, the branch has a well defined envelope.  b,e are of the chiral branching morphotype of the P. Dendritiformis bacteria.  Note that the bacteria of this morphotype are longer and have a quasi- 1D orientation.  c,f are of the P. Vortex bacteria.  Each dot on the macro-level corresponds to a vortex, one of which is shown on the micro-level.  Looking at the dynamics reveals that the bacteria are swimming in a coordinated manner around the common center of the vortex.  Comparing the different colonies, we note that as the macro-level complexity is elevated, the micro-level exhibits a higher degree of cooperation-based organization.

\item{Fig 2:} ÒBacterial CrystalsÓ Ð pattern formed during growth of the branching morphotype of the P. Dendritiformis bacteria when a 4-fold symetry mesh was ÒprintedÓ on the substrate.  The grooves shown in b have depths of less then 100 microns.

\item{Fig 3:} a. Simulation of the chiral branching structure using a walker model with an additional orientational degree of freedom. b. An example of spontaneous transition from the branching morphotype of P. Dendritiformis bacteria to the chiral branching one.  

\item{Fig 4:} Colonies grown without (left) and with (right) non-lethal antibiotic levels. Increase by a factor of 2 of the radius of curvature allows faster spreading. Bar is 5mm

\item{Fig 5:}  a) Rotating spiral wave in Xenopus occyte (courtesy of J. Lechleiter). b) Propagating wave as composed of individual ``puffs"  in oocyte experiments. Under these conditions the control run  exhibits global events whereas adding slowly-acting buffers (EGTA, in c.) eliminates the propagation (courtesy of I. Parker)

\item{Fig 6:} Confocal image to show mitochondria within a single cultured rat cortical astrocyte stained with the calcium-sensitive dye rhod-2 which partitions into mitochondria, permitting direct measurements of intramitochondrial calciuum concentration (courtesy, M. Duchen)Ê 

\end{trivlist}

\end{document}